\documentclass{iauJDSS}
\usepackage{graphicx}

\title[Helium spectrum in erupting prominences]
{The Helium spectrum \break in erupting solar prominences}

\author[Labrosse, Gouttebroze, Vial]
{N. Labrosse$^1$, P. Gouttebroze$^2$ \and J.-C. Vial$^2$}

\affiliation{$^1$Institute of Mathematical and Physical Sciences, \break University of Wales Aberystwyth, SY23 3BZ, UK \break email: nll@aber.ac.uk\\[\affilskip]
$^2$Institut d'Astrophysique Spatiale, CNRS -- Universit\'e Paris XI, \break 91405 Orsay cedex, France}

\pubyear{2006}
\volume{Volume 14}
\pagerange{119--126}
\date{?? and in revised form ??}
\jname{Highlights of Astronomy, Volume 14}
\editors{K.A. van der Hucht, ed.}

\newcommand\ion[2]{{#1}~\textsc{{#2}}}

\begin{document}

\maketitle

\begin{abstract}
Even quiescent solar prominences may become active and sometimes erupt. These events are occasionally linked to coronal mass ejections. However we know very little about the plasma properties during the activation and eruption processes. We present new computations of the helium line profiles emitted by an eruptive prominence. 
The prominence is modelled as a plane-parallel slab standing vertically above the solar surface and moving upward as a solid body. The helium spectrum is computed with a non local thermodynamic equilibrium radiative transfer code. The effect of Doppler dimming / brightening is investigated in the resonance lines of \ion{He}{i} and \ion{He}{ii} formed in the EUV, as well as on the \ion{He}{i} $\lambda\lambda$\,10830 \AA\ and 5876 \AA\ lines. We focus on the line profile properties and the resulting integrated intensities. 
It is shown that the helium lines are very sensitive to Doppler dimming effects.  We also study the effect of frequency redistribution in the formation mechanisms of the resonance lines and find that it is necessary to use partial redistribution in frequency for the resonance lines.
\keywords{Line: formation, Line: profiles, Radiative transfer - Sun: prominences}
\end{abstract}

\firstsection 

\section{Modelling of an eruptive prominence}

\begin{figure}
  \centering
  \resizebox{0.9\hsize}{!}{\includegraphics{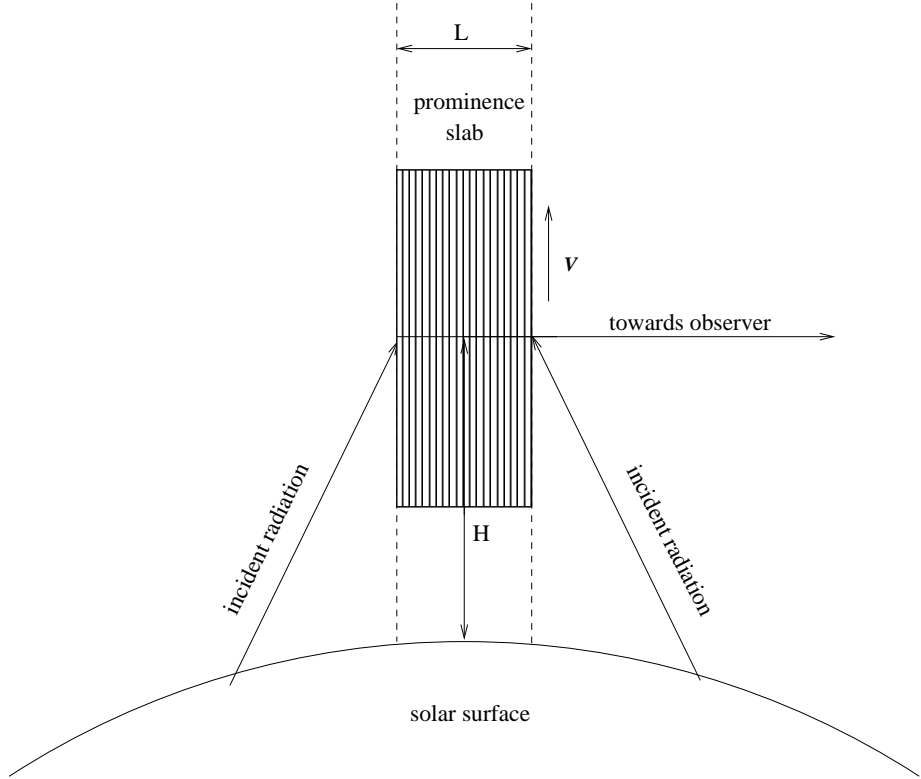}}
  \caption{Representation of our 1D plane-parallel prominence. The incident radiation coming from the solar disk depends on the plasma radial velocity $\vec{V}$ perpendicular to the line-of-sight.}
  \label{protu}
\end{figure}

We use our non local thermodynamic equilibrium radiative transfer code where the prominence is modelled as a plane-parallel slab standing vertically above the solar surface and moving upward as a solid body ({Fig.~\ref{protu}}). We consider isothermal and isobaric prominence atmospheres only. 
Other computational details are given in \cite{lgv06} (astro-ph/0608221).

\section{Frequency redistribution}

\begin{figure}
  \centering
  \resizebox{\hsize}{!}{\includegraphics{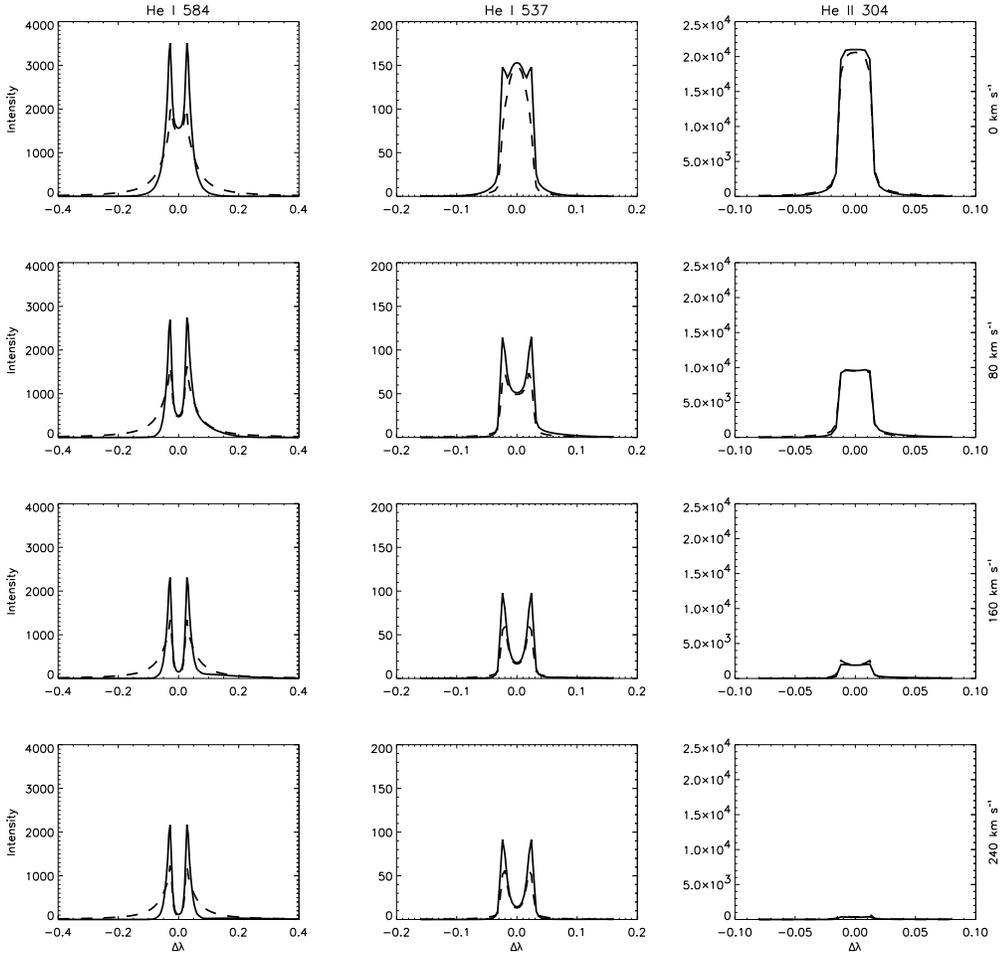}}
  \caption{Line profiles in CRD (dashed line) and PRD (solid line) for a prominence with height $H=50\,000$~km, width $L=650$~km, temperature $T=6500$~K, and pressure $p=0.1$~dyn cm$^{-2}$. The treatment of frequency redistribution is crucial for the first resonance line \ion{He}{i} $\lambda$\,584 \AA\ because of the predominant contribution of the line wings compared to the contribution of the Doppler core in the emitted intensity.}
  \label{profcrdprd}
\end{figure}

The redistribution in frequency during the scattering of the incident photons in resonance lines is best described by the so-called standard partial redistribution {(PRD)} approximation: a combination of complete frequency redistribution (CRD) and coherent scattering in the atom rest frame.
We compare the CRD case, where all helium lines are treated in CRD, with the PRD case, where PRD is used for the resonance lines \ion{He}{i} $\lambda\lambda$\,584 and 537~\AA\ and \ion{He}{ii} $\lambda$\,304~\AA. {Fig.~\ref{profcrdprd}} shows that there are substantial differences between the CRD and PRD treatments of the redistribution in frequency. {It is necessary to compute the helium spectrum in PRD} to compare the calculations with observations.

\section{Sensitivity to temperature}

We compute several prominence models at two different temperatures (8000~K and 15000~K) for velocities between 0 and 400~km~s$^{-1}$, keeping other parameters fixed, and using PRD. 

\begin{figure}
  \centering
  \resizebox{\hsize}{!}{\includegraphics{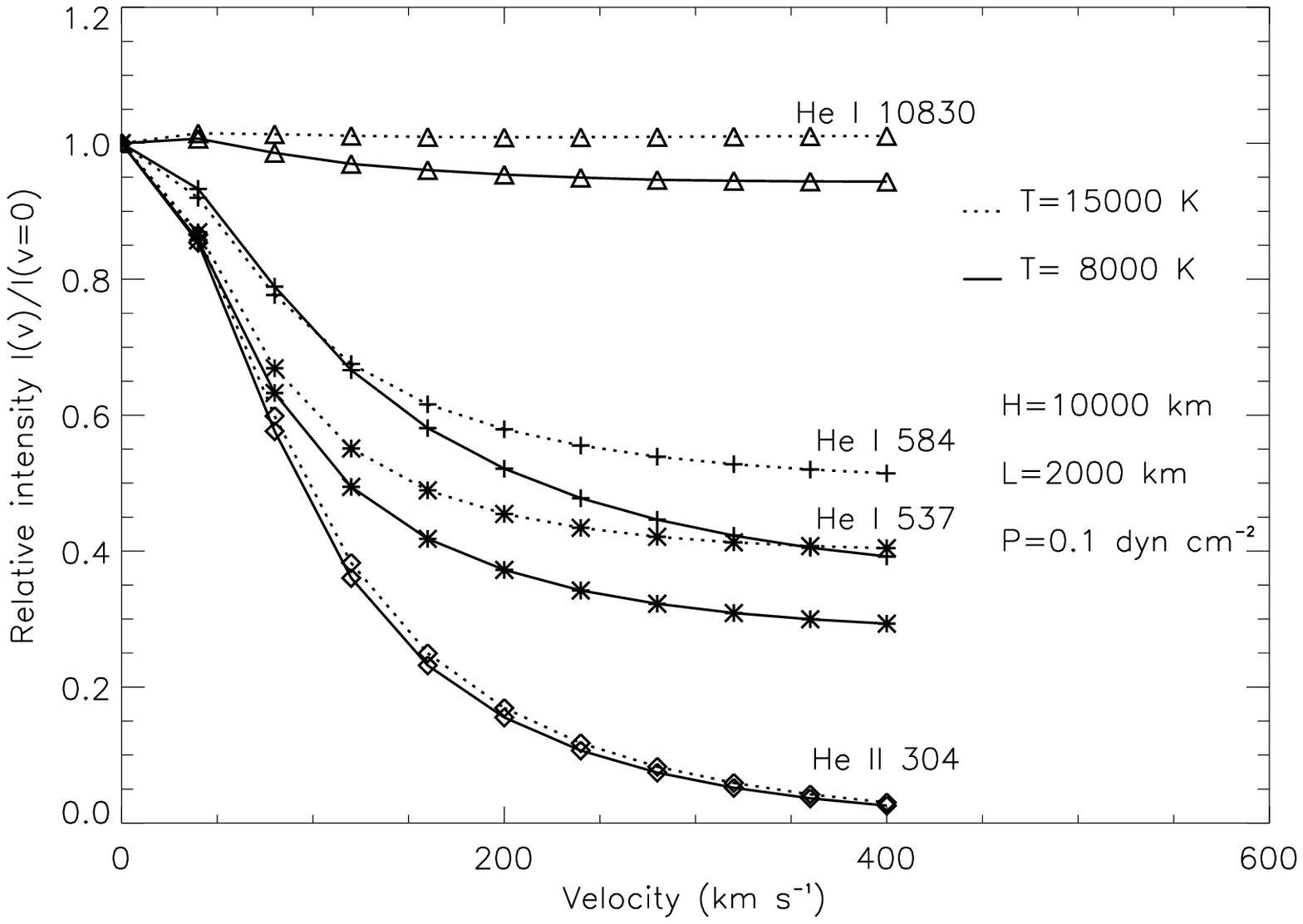}}
  \caption{Relative intensities (intensities normalized to the line intensity at rest) as a function of velocity at 8000~K and 15\,000~K for \ion{He}{i} $\lambda\lambda$\,10\,830, 584 and 537~\AA\ and \ion{He}{ii} $\lambda$\,304~\AA. The \ion{He}{ii} line is the most sensitive to the velocity of the erupting prominence.}
  \label{tempint}
\end{figure}

Fig.~\ref{tempint} shows that there is little difference between the two temperatures for the relative intensity of the \ion{He}{ii} line. This line is indeed mainly formed by resonant scattering of the incident radiation, as shown in \cite{lg01}. 
The \ion{He}{i} resonance lines show some sensitivity to the temperature: at high temperatures,  collisional excitation becomes non negligible. The difference between the cold and hot prominences is more evident as the  velocity increases. The  contribution of resonant scattering in line formation decreases with increasing velocity. Beyond about 200~km s$^{-1}$, the thermal contribution dominates the emission. Therefore the emergent intensity becomes almost independent of the velocity. 
The \ion{He}{i} 10830 line does not strongly depend on  velocity. The incident radiation coming from the solar disk at this wavelength is a very weak absorption line. Consequently, the Doppler effect is virtually unseen in the emergent line. Finally, \ion{He}{i} D3 (5876 \AA, not shown) behaves exactly as \ion{He}{i} 10830.

\begin{figure}
  \centering
  \resizebox{\hsize}{!}{\includegraphics{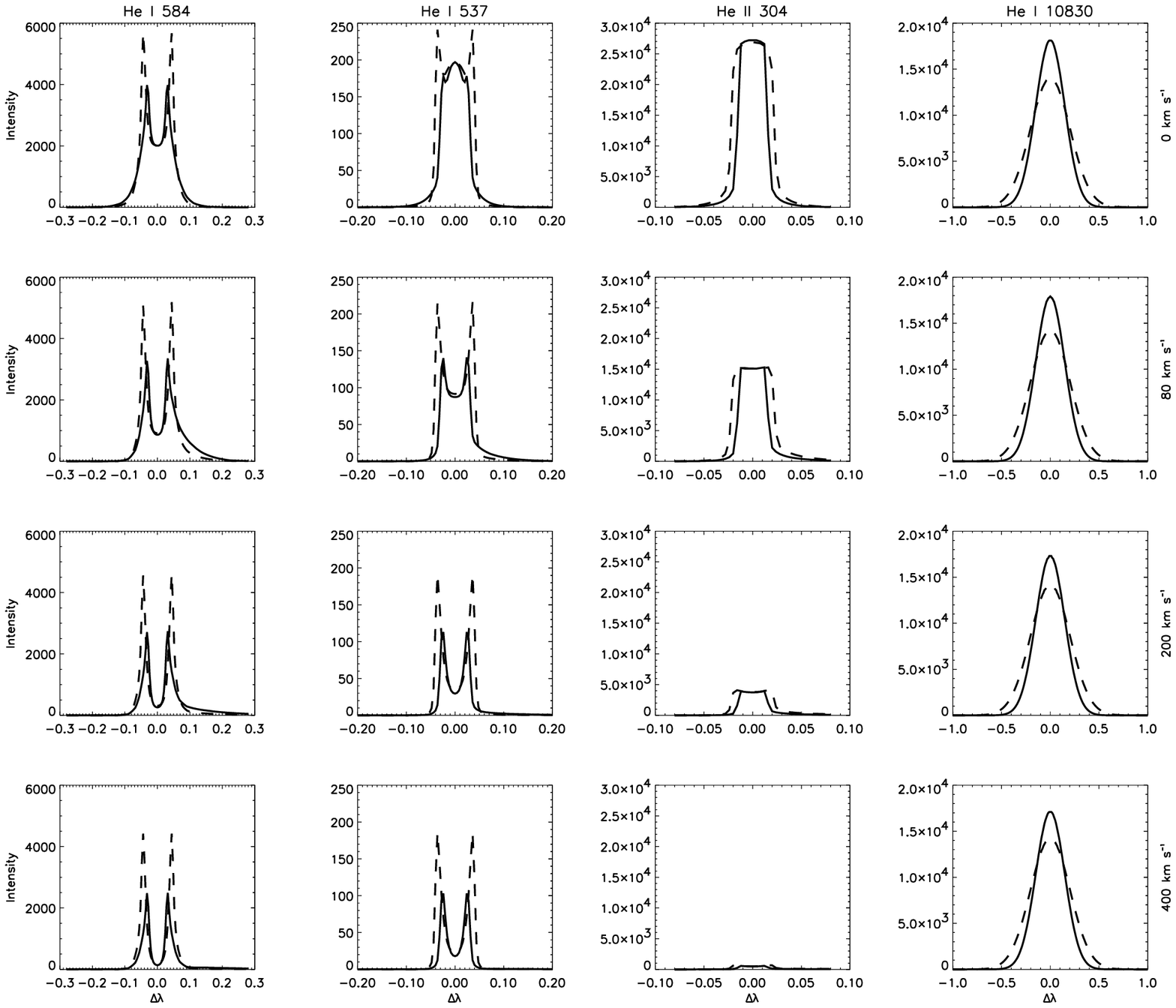}}
  \caption{Line profiles of \ion{He}{i} $\lambda\lambda$\,584 \AA, 537 \AA, \ion{He}{ii} $\lambda$\,304 \AA, and \ion{He}{i} $\lambda$\,10\,830 \AA\ for $T=8000$~K (solid line) and $T=15\,000$~K (dashed line), $p=0.1$~dyn~cm$^{-2}$, and $L=2000$~km. The first row corresponds to the case where the prominence is at rest, and the line profiles are symmetrical. Then the velocity increases from 80 km s$^{-1}$ (2nd row) to 200 km s$^{-1}$ (last row). The Doppler dimming effect is clear at all temperatures for the three resonance lines, but not the 10\,830 \AA\ line.}
  \label{proft}
\end{figure}

Figure~\ref{proft} illustrates how the line profiles are affected by the velocity at two different temperatures.

 \subsection{\ion{He}{i} $\lambda$\,584~\AA\ and \ion{He}{i} $\lambda$\,537~\AA}

At low temperature and low speeds, we observe an asymmetry in the line profiles of the two resonance lines, with some intensity enhancement in the red part. 
At high speeds, a smaller asymmetry still exists.
At high temperature, the asymmetry in the line profile is less pronounced than at low temperature. 
At a given temperature, the red wing of the profile first increases, and then decreases, with increasing velocity.
The asymmetries due to the enhanced emission in the red wings are more visible at speeds below 200~km s$^{-1}$ when thermal emission is negligible.
Finally, the reversal at line centre is more pronounced with increasing velocity.
  
\subsection{\ion{He}{i} $\lambda$\,10,830~\AA}

  The intensity in this line does not strongly depend on the plasma velocity, due to the very weak incident absorption line.

\subsection{\ion{He}{ii} $\lambda$\,304~\AA}

The emitted intensity in the \ion{He}{ii} line mainly depends on the incident radiation.
At a given speed, the intensity at line centre is the same for both temperatures. 
The line profile becomes broader with temperature.
It is easily seen from Figs.~\ref{tempint} and \ref{proft} that the 304 \AA\ line intensity  decreases rapidly with increasing velocity.

  \section{Conclusions}

    {Partial redistribution in frequency is necessary} to compute the profiles of the helium resonance lines emitted by moving material in prominences. 
    {Velocity effects are more visible when thermal emission is low.}
Together with hydrogen lines the helium lines offer the possibility of a powerful diagnostic of the eruptive prominence plasma.
In a future work, we will infer {the full velocity vector} by combining {Doppler dimming / brightening effects} on hydrogen and helium lines with the {apparent motion} of the prominence material brought by SOHO or future imagers (e.g. SOLAR-B, STEREO).

\begin{acknowledgments}
NL acknowledges financial support from PPARC grant PPA/G/O/2003/00017, and IAU grant 12624 to the XXVI General Assembly of the IAU.
\end{acknowledgments}

\end{document}